\documentstyle{article}
\newlength{\figwidth}
\newlength{\refspace}

\begin{document}
\newcommand{\be}{\begin{equation}}
\newcommand{\ee}{\end{equation}}
\newcommand{\bra}{\langle}
\newcommand{\ket}{\rangle}
\newcommand{\del}{\partial}
\renewcommand{\arraystretch}{1.8}
\setcounter{equation}{0}
\def\real{{\rm I\kern-.2em R}}
\def\complex{\kern.1em{\raise.47ex\hbox{
            $\scriptscriptstyle |$}}\kern-.40em{\rm C}}
\def\integer{{\rm Z\kern-.32em Z}}
\def\half{\frac{1}{2}}
\def\ra{\rightarrow}
\title{Piezoelectricity: Quantized Charge Transport Driven by Adiabatic
Deformations}
\author{J.E.~Avron ${}^{a*}$, J.~Berger$^a$ and Y.~Last$^b$\\
$^a$ Department of Physics, Technion, 32000 Haifa, Israel\\
$^b$ Division of Physics Mathematics and Astronomy, Caltech, Pasadena,
CA 91125\\
\it{ \footnotesize ${}^*$  e-mail avron@physics.technion.ac.il}}
\maketitle

\begin{center}
{\bf Abstract}
 \end{center}
{\em We study the (zero temperature) quantum piezoelectric response of
Harper-like
models with broken inversion symmetry. The charge transport in these
models is related to topological invariants (Chern numbers). We show
that there are arbitrarily small periodic modulations of the
atomic positions that lead to nonzero charge transport for the
electrons.}

The Harper model can be interpreted as a tight-binding quantum
Hamiltonian describing the dynamics of non-interacting electrons on
a two dimensional lattice in the presence of magnetic fields. It is
known to have interesting Hall transport properties. Here we study
the electric response of Harper-like models to adiabatic changes in 
the hopping amplitudes. Changes in the hopping amplitudes have a
natural interpretation as elastic deformation of the underlying lattice.
As we shall show, such deformations can drive electron transport. We
shall refer to this kind of response as piezoelectricity. Like the
Hall conductance in the integer Hall effect
\cite{tknn,stone}, and  in quasi-one dimensional systems
\cite{montambaux},  the Thouless pump \cite{thouless,niu},
the Magnus force
\cite{magnus}, adiabatic charge transport in networks \cite{arz},
adiabatic spin transport
\cite{haldane}, and adiabatic viscosity \cite{asz}, it is a transport
phenomenon related to the adiabatic curvature and Berry's phases
\cite{Berry}.

Let us  first summarize the central findings: 1.\ Harper-like models 
with broken time reversal and broken inversion symmetry have, in 
general, nontrivial piezoelectric response. 2.\ Appropriate
{\em periodic} modulations of the atomic positions lead to integral
charge transport given by appropriate Chern integers.
This implies that an ac driving  has a response with a dc component.
3.\ There are {\em arbitrarily small} periodic deformations that
transport integral (and nonzero) charges over macroscopic distances.
These periodic cycles trap level crossings in parameter space.

The topological significance of piezoelectricity was noted in
\cite{asz}, but vanished for the models considered there
(Landau Hamiltonians). Thouless \cite{thouless} and subsequently 
Niu \cite{niu} constructed one dimensional models of charge pumps
where integral charge transport (given by a Chern number) is driven
by one periodic potential sliding past another. Here, the model,
the driving and the details are different, but the spirit is the same.

We shall focus on a family of Harper models,
$H(\vec t,\vec k,\eta,\phi)$,
which arises from tight binding models associated with a two
dimensional  triangular lattice. Each site of the lattice has a
coordination number six and the basic plaquettes are triangles. Each
up-triangle in the lattice is surrounded by three down-neighbors and
vice
versa. The magnetic flux through the up-triangles is $\phi/2+\eta$ and
$\phi/2-\eta$ through the down-triangles. $\phi=2\pi\ p/q$, with $p,q$
relative primes, and $|\eta|\le \pi/2$  is a measure of the asymmetry in
the fluxes through the up/down triangles in units where the quantum of
flux is $2\pi$.  The hopping amplitudes associated with the three basic
vectors of the triangular lattice are
$t_j\in\real$, $j=1,2,3$. The corresponding  Harper model is:
 \begin{eqnarray} \left(H(\vec t,\vec k,\eta,\phi)\Psi\right)(n) =
 \left(t_1+ t_3y_ne^{i\phi/2}\right) x \Psi({n}+1)\ +\nonumber\\
2 t_2\, \cos(n\phi +k_2)\Psi( n)+\left(t_1+t_3\bar y_n
e^{i\phi/2}\right) \bar x \Psi(n-1).\label{matrix}\end{eqnarray}
  $x=\exp (i k_1),\ y_n=\exp i(n\phi+k_2 -\eta),\
\Psi(n+q)=\Psi(n)\in\complex$ and $\bar x,\bar y_n$ are the complex
conjugates of $x,y_n$. $\vec k$ are Bloch momenta with ranges $|k_1|\le
\pi/q$, $|k_2|\le \pi$. The model was introduced
in \cite{bellisard} who studied the Hofstadter spectrum in the case
$\vec t =(1,1,1)$.

The class of models in
Eq.~(\ref{matrix}) is the simplest among Harper-like models with
interesting piezoelectric response. The simpler versions of the Harper
model and, in particular, the classical Harper model on the rectangular
lattice  and its generalizations
\cite{han}, do not have interesting adiabatic piezoelectric
response.  The reason for this is that  inversion symmetry needs to be
broken. This is a fact about piezoelectricity that goes back to
the brothers Curie
\cite{nye}. Inversion
symmetry is broken if
$\eta\neq 0  \;{\rm mod}\, \pi$. Inversion symmetry is preserved in the
classical Harper model and the generalizations studied in \cite{han}.

Let $|\psi(\vec t,\,\vec k,\,\eta)\rangle$ be a normalized Bloch
state of the Harper model in Eq.~(\ref{matrix}). Consider a closed
loop $\gamma\subset\real^3$ in the space of hopping amplitudes.
When $\gamma$ is traversed
adiabatically (this, of course, subsumes that the gap remains open),
the charge $Q(\gamma, k_2,\eta)$ transported from
$-\infty$ to $\infty$  in the $k_1$ direction for fixed $k_2$ channel
for
each full band is given by
\cite{thouless,arz}
\begin{eqnarray}
Q(\gamma, k_2,\eta)=\frac{1}{\pi}{\rm Im}\,  \int_{-\pi/q}^{\pi/q}\
dk_1\
\int_\gamma
\langle\frac{\partial\psi}{\partial k_1}|\vec\nabla_t\psi\rangle\cdot
d\vec t.
\label{chern}\end{eqnarray} The charge, if well defined, is an
integer---a Chern number. The total charge transported by the system is
the
sum over the  relevant $k_2$ channels  and the occupied bands. When the
system is an infinite two dimensional crystal then all the $k_2$
channels
are relevant. On the other hand, for a strip of finite width with
(possibly
twisted) periodic boundary conditions, only a discrete set of values of
$k_2$ contributes. For reasons that shall become clear later, finite
strips
are the more interesting case.

The difficulties in studying Chern numbers of model Hamiltonians
\cite{tknn,arz,asz,novikov} (and this one is no exception) are: First,
one needs to establish that the Chern numbers are well defined. For the
problem at hand, this means that one needs to isolate a range of
parameters
$k_2, \eta$ and $\phi$ for which the gaps surrounding an energy band
remain open when
$\vec t$ and $k_1$ run over their full range. Second,
the Chern number may
be well defined but zero, a case that is not very interesting for
transport.
For this not to be the case, the surface of integration in
Eq.~(\ref{chern})
must  be protected against contraction.
For certain transport properties
such as those considered, e.g., in
\cite{tknn,arz,haldane}, the surface of integration
had such a protection
built in. This is not the case here.
The cycle of deformations,
$\gamma$, is a closed orbit in the three dimensional space of
deformations, and such an orbit can be contracted to a point. If during
this contraction  the integrand in Eq.~(\ref{chern}) remains continuous,
the Chern number is zero. So, for the Chern number to be nonzero, the
orbit
of deformations $\gamma$ {\em must trap level crossing}. Finally, one
needs
to worry about global questions:
$Q(\gamma,k_2,\eta)$ must be well defined for {\em all} of the relevant
$k_2$ channels and must not sum up to zero. It turns out that the Harper
model is rich enough so that everything actually happens there; there
are
good orbits and parameters  where one finds nonzero quantized transport,
but also bad ones where various bad things happen.

The Bloch Hamiltonian, Eq.~(\ref{matrix}), is a homogeneous function of
$\vec t$ of order one:
$H(\vec t,\vec k,\eta ,\phi) = |\vec t|\  H(\hat t,\vec k,\eta, \phi )$.
The eigenvectors are independent of
$|\vec t|$, and the length of $\vec t$ therefore does not contribute to
Eq.~(\ref{chern}).   We shall henceforth take $\hat t$ to be on the unit
sphere. The  three additional continuous parameters, $\vec k$ and $\eta$
are angular variables. Eq.~(\ref{matrix}) depends on five continuous
parameters,
$(\hat t,\vec k,\eta)$. The five dimensional parameter space is
topologically the product of a two-sphere and a three-torus.

To get one's hands on the Chern numbers for this model, one needs, as we
have seen, to have good control over level crossings. One can use
symmetry considerations to reduce the study of crossings from the full
range of the parameters to a part of the parameter space. Indeed, there
are
three linear transformations of the parameters which are implemented by
either unitary or anti-unitary transformations. These are
\begin{eqnarray}\{ k_j \ra k_j+2\pi/q\},\quad
\{ \eta \rightarrow -\eta,\ \vec{k} \rightarrow -\vec{k}\}, \nonumber\\
 \{ k_1 \rightarrow \eta-k_1-k_2+\left(1-(-1)^q\right) \phi /4,
\ t_1\leftrightarrow t_3 \}\label{map}\end{eqnarray}
 As a consequence of this, the spectral analysis of $H(\hat t,\vec
k,\eta
,\phi)$,  can be restricted to the range:
\begin{eqnarray} -\pi/q\le k_j <\pi/q, \quad 0\le \eta<\pi/2, 
\quad t_3\le t_1. \label{range}\end{eqnarray}
We shall take $t_1$ and $t_2$ to be our coordinates on the sphere of
deformations. For the sake of concreteness we  restrict
ourselves to the positive quadrant $t_j\ge 0,\ j=1,2,3$,
and to the ground
state of Eq.~(\ref{matrix}). We shall call the point on the unit sphere
with $t_j=1$ ``the j-th pole.''

Let $\Gamma$ be the set of points where the lowest eigenvalue of
Eq.~(\ref{matrix}) is degenerate. By $\Gamma(k_2=c)$ we shall denote the
restriction of $\Gamma$ to the subspace with fixed channel $k_2=c$
and by $\Gamma(k_2=c,\eta=d)$ we denote the restriction to a fixed
channel and asymmetry, etc. Recall that the  von Neumann-Wigner rule
\cite{wigner} says that a complex Hermitian matrix which depends on $d$
parameters has, generically, eigenvalue crossings on a surface of
dimensions
$d-3$. One therefore expects
$\Gamma$ to be two dimensional surfaces, $\Gamma(k_2=c)$ to be   one
dimensional curves and $\Gamma(k_2=c,\eta=d)$ to be isolated points. We
shall see that this is a good guide  to the behavior
of the set of level crossings away from special points, e.g., the poles.
For a generic point of
$\hat t$, the von Neumann-Wigner rule says that
$\Gamma(\hat t)$ is a discrete set of points in $\vec k \otimes \eta$ 
space.
At the poles  we shall find,  instead, that $\Gamma(t_j=1)$ is a two
dimensional surface. Of course, the poles are special points, and the
failure of  von Neumann-Wigner there is no source of concern.

At the poles Eq.~(\ref{matrix}) can be diagonalized by hand.
At the 2-pole
the Hamiltonian is already in a diagonal form.  At the 1-pole  it is
diagonalized by plane waves and at the 3-pole, by plane waves up to an
appropriate gauge transformation.   The restrictions of $\Gamma$ to the
poles,
$\Gamma(t_j=1)$,  can be determined explicitly. More precisely,
$\Gamma(t_j=1)$  is the  2D set of points that obey:
\begin{eqnarray} 
 k_1&=&(1+(-1)^q)\phi /4, \qquad\qquad\qquad\, {\rm for\ } j=1;
\nonumber\\
 k_2&=&-(1+(-1)^q) \phi/4, \qquad\qquad\quad\;\,  {\rm for\ }
j=2; \label{poles}\\
 k_1&=&- k_2 + \eta\ +
\left(1-(-1)^q \right)\phi /4, \quad {\rm for\ } j=3. \nonumber
\end{eqnarray} 
The degeneracies at the 1-pole and the 3-pole are related by
symmetry, Eq.~(\ref{map}).

Let us now consider the special cases $q=1,2,3$: The case $q=1$
corresponds
to $\phi=0$ and  is trivial; the Bloch Hamiltonian has one eigenvalue,
no crossing, and no charge transport.  The case $q=2$
(or equivalently, $\phi=\pi$) is already  interesting.  The Bloch
Hamiltonian,  Eq.~(\ref{matrix}) reduces to the basic paradigm for
Chern numbers---Berry spin 1/2 Hamiltonian:
\begin{eqnarray}
\left(t_2 \cos k_2\right) \sigma_3+
\left(t_1 \cos k_1 \right) \sigma_1+
\left(t_3 \cos(k_1+k_2-\eta)\right) \sigma_2,\label{pauli}
\end{eqnarray}
with $\sigma_j$ the Pauli matrices.
Since the matrix is traceless, levels cross when it vanishes.
This gives Eq.~(\ref{poles}) and is all of
$\Gamma(k_j)$, provided $k_j\neq\pi/2$. At these special points
$\Gamma(k_2=\pi/2)$  is the two great circles $t_1=0$ and $t_3=0$;
similarly,  $\Gamma(k_1=\pi/2)$  is the two great circles $t_2=0$ and
$t_3=0$. If $k_1=k_2=\eta=\pi/2$, then the whole unit sphere
$|\hat t|=1$ belongs to $\Gamma$.

Now that the set of level crossings is known, we can describe the Chern
numbers. By the general principles mentioned before,
interesting Chern numbers arise  when the orbit in deformation space
$\gamma$ traps  level crossings.  Let
$\gamma_j$ denote a small  closed orbit around the j-th pole.
For  $k_2\neq\pi/2$ these orbits trap level crossings and are such that
the Chern number, Eq.~(\ref{chern}), is well defined. The charge
transport can be computed by a formula of \cite{simon,arz}:
\begin{eqnarray}
Q(\gamma_1,k_2,\eta) &=& \mp {\rm sgn}(\cos k_2)\,
{\rm sgn}(\sin(k_2-\eta)),\nonumber\\
Q(\gamma_2,k_2,\eta) &=&0,\label{signs}\\
Q(\gamma_3,k_2,\eta) &=& \pm {\rm sgn}(\cos k_2)\, {\rm sgn}
(\sin\eta).\nonumber
\end{eqnarray} The overall sign depends on the orientation of
$\gamma_j$, and is opposite for the top/bottom bands.
The Chern numbers change (discontinuously) on $\Gamma$ so the
direction of charge transport can be flipped by tuning
$k_2$ and $\eta$.

For $k_2=\pi/2$, the
Chern number $Q(\gamma_2,k_2,\eta)$ is not well defined since there are
level crossings on the surface of integration.  The {\em total} charge
transport is a well defined integer if $k_2=\pi/2$ is not an allowed
channel, and is ill defined  if this channel is allowed. For finite
strips {\em with periodic boundary conditions}, odd strips  have
$k_2\neq\pi/2$ and the total charge transport is integral. Even strips,
and also the infinite two dimensional lattice, include the $k_2=\pi/2$
channel and do not have a well defined (total) Chern number. For finite
strips where the channel
$k_2=\pi/2$ is excluded, the total transport can be read off from
Eq.~(\ref{signs}). In particular, with maximal breaking of inversion
symmetry,
$\eta=\pi/2$, an orbit of deformations $\gamma_1$ about the 1-pole,
transports $\# \{k_2\ {\rm channels}\} $ charges in the ground state.
The total charge transport is therefore a nonzero integer for any
strip (where the number of $k_2$ channels is finite) and can be
arbitrarily large. This shows that summation over the
$k_2$ channels does not cancel in general: in this case, they add.
In contrast, for an orbit of deformations
$\gamma_3$ around the 3-pole, the total charge transport is $\pm 1$ for
all
odd strips.  This is because the allowed values of
$k_2$ are equally spaced and then $\sum {\rm sgn} (\cos k_2) = \pm 1.$

We see
from this that: 1.\ The Harper model, Eq.~(\ref{matrix}), has nontrivial
piezoelectric response. 2.\ For appropriate values of parameters and
orbits,
the charge transport is given by nonzero Chern integers. 3.\ The Chern
numbers can sum to nonzero integers when summation over channels is
taken.
4.\ Integer transport occurs also for arbitrarily small deformations
$\gamma$.

One may criticize the $q=2$ example of piezoelectric response as
being too special in that the deformations $\gamma$ that give charge
transport are about  points in parameter space where two hopping
amplitudes vanish. This is a rare event, analogous to
multicriticality. Can one have piezoelectric transport also if all
hopping amplitudes remain positive?
As we shall see, this happens for the Harper model with $q=3$.
The price we
shall pay is that the analysis of the set of level crossing is more
complicated and part of it relies on detailed numerical analysis.

For $q=3$, the model is described by a $3 \times 3$ matrix with the
characteristic polynomial
\begin{eqnarray}
E^3&-&3E=2h(\hat t,\vec k, \eta)=t_1^3 \cos 3k_1+\label{cp}\\&t_2^3&
\cos 3k_2 - t_3^3\cos\left(3(k_1+k_2 - \eta) \right)+
3t_1t_2t_3\cos(\eta). \nonumber
\end{eqnarray}
Eq.~(\ref{cp}) is a strong version of Chamber's
relation: the coefficients of $E$ are not only independent of $\vec{k}$,
but also of
$\hat{t}$ and $\eta$. Therefore, the band edges are at extrema of $h$ in
the entire five-dimensional parameter space. The set of curves where the
first
gap closes for $q=3$ is obtained when $E=-1$ and
$h=1$.

The strategy we use to get hold of the degeneracy surface $\Gamma$ is
the following: At the 1-pole, Eq.~(\ref{poles}) gives a two dimensional
planar piece of $\Gamma$.
(This two dimensional plane projects to a line in Fig.~1.) The
line
$\Gamma(t_1=1,\eta =0)$ turns out to be a line of self-intersection of
$\Gamma$. One two dimensional piece is given in Eq.~(\ref{poles}). The
intersecting two dimensional piece can be obtained as follows.
Pick any point on $\Gamma(t_1=1,\eta =0)$   and expand
$h$ in powers of
$t_2$ and require that $h=1$ to every order.
If we now use $k_2$ and $t_2$ as
the parametric representation of $\Gamma$ we find:
\begin{eqnarray} k_1&=&\sin(3 k_2) t_2^3/3+\sin(3 k_2)
t_2^5+...\quad ;\nonumber\\
 t_1&=&1-t_2^2+\cos(3 k_2) t_2^3+...\quad ; \label{10}\\
\eta&=&-\sin(3 k_2) t_2-\sin(6 k_2) t_2^2/2+...\quad .\nonumber
\end{eqnarray}
In this parametric
representation, $t_2$ is small and $k_2$ is arbitrary.
This gives us a thin
strip of $\Gamma$, which intersects that of Eq.~(\ref{poles}). We can
extend this strip using the fact that the tangent plane to
$\Gamma$ is the kernel of the Hessian of $h$. In other words, with $k_2$
fixed, the curve of degeneracy, $\Gamma(k_2)$, may be described as the
solution of an ordinary differential equation: The velocity in parameter
space is given by the kernel of the Hessian of $h$. Near the 3-pole,
$\Gamma(k_2)$ is given by
 (\ref{10}) and (\ref{map}).

Several curves describing degeneracies are shown in the figure. We chose
$t_1$ and $t_2$ as our coordinates on
$\hat t$ restricted to the positive quadrant,
$0\le t_j\le 1.$  Let us denote by $\Gamma_c$ the degeneracy surface
with the poles excised.
The tongue-like curves in Fig.~1 are
$\Gamma_c(k_2=-\pi/18)$ and
$\Gamma_c(k_2=-\pi/3)$. By the
von Neumann-Wigner rule one expects these to be one dimensional curves,
and indeed they are. For
the orbit of deformation  $\gamma$  shown in this figure, that is,
a small circle centered at $t_1=0.355, t_2=0.446$, the charge transport
is $Q(\gamma, -\pi/18,\pi/9)=\pm 1$ (the sign depends on the orientation
on which $\gamma$ is traversed). This gives an example where a Chern
number is nonzero for a small orbit of deformations that lies entirely
in the positive quadrant of hopping amplitudes.

For $q=2$, we have seen that Chern numbers for the infinite crystal
included channels with ill defined Chern numbers. One may wonder
if this is also the case for $q=3$. The answer is no. For $q=3$, all
sufficiently small paths
$\gamma_j$ around the poles have well defined Chern numbers for {\em
all}
$k_2$ channels, and some of these are nonzero.  It is easy to verify
this for $\eta=\pi/2$ where one can check that all of $\Gamma$ is at the
poles.  More is true; $\Gamma$ is in fact restricted to the poles for
all
$\pi/6 <\eta\le\pi/2$. One way to see this, is by analysis at the
vicinity of the point
\hbox{$\hat t=1/\sqrt{3}(1,1,1)$}, $\vec k=-\pi/18(1,1),
\ \eta=\pi/6$. One finds that
$\eta$ attains its  maximum value on $\Gamma_c$ at this point.
Further study shows that, in fact, for all $\eta\neq 0$ mod $\pi$ a
sufficiently small orbit $\gamma_j$ about the j-pole avoids $\Gamma$.
The Chern numbers for these orbits
$\gamma_j$ are all well defined, and by numerical integration we found
$Q(\gamma_j,k_2,\eta)= \pm 1, 0,\mp 1$ for   $j=1,2,3$ pole
respectively, for all $k_2$ channels and $\eta\neq 0$.

In conclusion: we have described a method for analyzing the Chern
numbers that arise in inversion asymmetric Harper models and have
found explicit situations with nonzero quantized piezoelectric
response. In all these cases nonzero transport occurs for arbitrarily
small orbits, and this can happen also when all the hopping amplitudes
are positive.

We thank A.~Auerbach and P.~Zograf for discussions and E.~Akkermans for
bringing ref. \cite{montambaux} to our attention. This work is supported
in part by GIF, the Israel Science Foundation, the DFG and by the
Fund for the Promotion of Research at the Technion. YL acknowledges
the hospitality of ITP at the Technion and JA the hospitality of
Caltech.

\newpage
{\bf Figure Caption:}

Curves on which  the first gap closes for $q=3$. The vertical
axis
is the asymmetry flux $\eta$. The horizontal plane is the  positive
quadrant in the plane
$0\le t_1,t_2\le 1, \ \ t_1^2+t_2^2\le 1$. The hatched region is this
positive quadrant at $\eta=\pi/9$. The two vertical lines at $t_2=0$
correspond  to the
gap closure in Eq.~(\protect{\ref{poles}}). The tongue-like curve is the
line of  gap closure at $k_2=-\pi/18$.  It links with a small circle
$\gamma$ in the hatched plane, centered  at
$(t_1=.355, t_2=.446)$. A periodic deformation along $\gamma$ transports
a unit of charge. The curve in the plane $\eta=0$ is the line of gap
closure for $k_2=-\pi/3$.
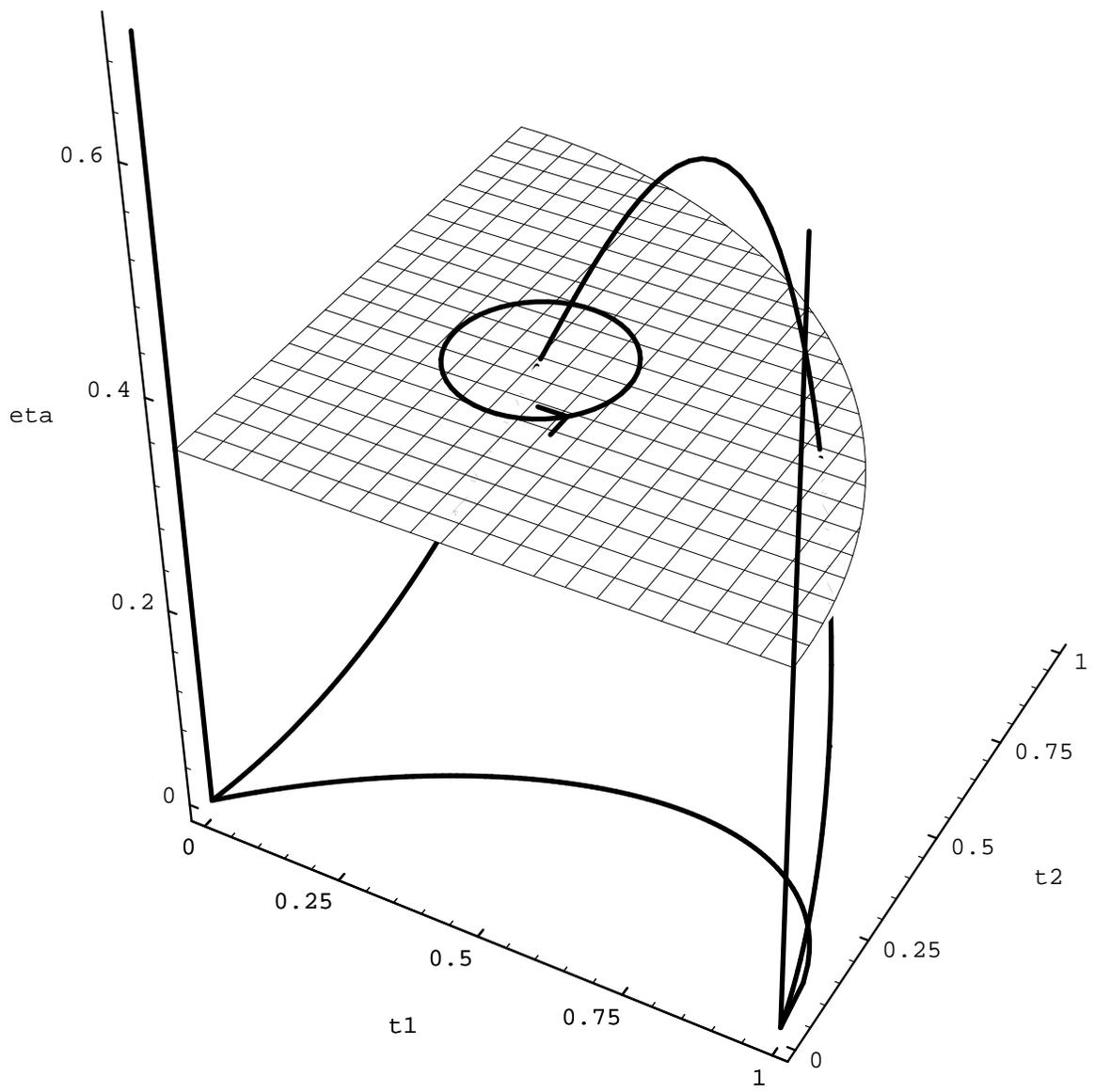
\end{document}